\documentclass[12pt]{iopart}
\usepackage{iopams}  

\usepackage{graphicx}
\usepackage{subfig}
\usepackage{bm}

\newcommand{\be}{\begin{equation}}
\newcommand{\ee}{\end{equation}}

\newcommand{\lk}{\left(}
\newcommand{\rk}{\right)}
\newcommand{\lkk}{\left\{ }
\newcommand{\rkk}{\right\} }

\newcommand{\R}{\mathbb{R}}

\begin{document}

\title[Exact calculation of the mean first-passage time of 
continuous-time random walks]
{Exact calculation of the mean first-passage time of 
continuous-time random walks
by nonhomogeneous Wiener--Hopf integral equations 
}

\author{M Dahlenburg$^{1,2}$, G Pagnini$^{1,3}$}

\address{$^1$ BCAM--Basque Center for Applied Mathematics,
Alameda de Mazarredo 14, 48009 Bilbao, Basque Country -- Spain}
\address{$^2$ Institute for Physics \& Astronomy, 
University of Potsdam, 14476 Potsdam, Germany}
\address{$^3$ Ikerbasque--Basque Foundation for Science,
Plaza Euskadi 5, 48009 Bilbao, Basque Country -- Spain}

\ead{mdahlenburg@bcamath.org}

\vspace{10pt}
\begin{indented}
\item[]Received 17 January 2022
\item[]Accepted 12 December 2022
\end{indented}

\begin{abstract}
We study the mean first-passage time (MFPT) 
for asymmetric continuous-time random walks in continuous-space
characterised by waiting-times with finite mean
and by jump-sizes with both finite mean and finite variance.
In the asymptotic limit, 
this well-controlled 
process is governed by an advection-diffusion equation and 
the MFPT results to be finite when the advecting velocity is in
the direction of the boundary. 
We derive a nonhomogeneous Wiener--Hopf integral equation  
that allows for the exact calculation of the MFPT 
by avoiding asymptotic limits 
and it emerges to depend on the whole distribution of 
the jump-sizes and on the mean-value only of the waiting-times,
thus it holds for general non-Markovian random walks.
Through the case study of a quite general family 
of asymmetric distributions of the jump-sizes 
that is exponential towards the boundary 
and arbitrary in the opposite direction,
we show  
that the MFPT is indeed independent of the jump-sizes distribution 
in the opposite direction to the boundary.
Moreover, 
we show also that there exists a length-scale, 
which depends only on the features of the distribution 
of jumps in the direction of the boundary,
such that 
for starting points near the boundary the MFPT depends 
on the specific whole distribution of jump-sizes, 
in opposition to the universality emerging 
for starting points far-away from the boundary. 
\end{abstract}

%
\noindent{\it Keywords}: 
mean first-passage time, 
continuous-time random walk, Wiener--Hopf integral equation

\submitto{J. Phys. A: Math. Theor.} 
%
\maketitle
%
%

\section{Introduction}
The mean first-passage time (MFPT) statistics are important,
in general, for many diffusive processes 
\cite{bray_etal-ap-2013,
metzler_etal-firstpassagetime-book-2014,
jpaspecialissue-2020}
and, in particular,
for living 
\cite{bressloff_etal-2014,chou_etal-2014,polizzi_etal-ijc-2016}
and reacting chemical systems 
\cite{weiss-acp-1967,szabo_etal-jcp-1980,hanggi_etal-rmp-1990}.
A large number of results have been derived 
\cite{redner-2001,metzler_etal-firstpassagetime-book-2014,
nyberg_etal-njp-2016}, 
together with recent findings both for Brownian 
\cite{metzler-jsm-2019,
hartich_etal-jpa-2019, 
majumdar_etal-jsm-2020,
grebenkov_etal-njp-2020,
kearney_etal-jpa-2021,
eliazar-jpa-2021}
and non-Brownian motions as 
the diffusing-diffusivity approach 
\cite{lanoiselee2018diffusion,sposini_etal-jpa-2019,grebenkov_etal-jpa-2021}
or L\'evy-like motions  
\cite{majumdar_etal-jpa-2017,padash_etal-jpa-2019,
padash_etal-jpa-2020,palyulin_etal-njp-2019}.
Beside established findings, 
the quantitative experimental analysis of first-passage time distributions
is becoming accessible in biology only recently
\cite{thorneywork_etal-sa-2020,broadwater_etal-bj-2021}
and further advancements
are still of high interest because of new emerging applications 
and results
\cite{
hartich_etal-njp-2018,
grebenkov_etal-njp-2020b,
grebenkov_etal-njp-2021a,
grebenkov2021distribution,
simpson_etal-njp-2021}.

Here, 
we focus on the MFPT problem in the well-controlled setting of a 
one-dimensional 
continuous-time random walk (CTRW) \cite{montroll_etal-jmp-1965}
in a continuous-space
when the waiting-times distribution has a finite mean
and the jump-sizes distribution has both finite mean and 
finite variance.
If the mean value of the jump-sizes is different from zero, 
in the asymptotic limit, 
the probability density function (PDF) of such random walker 
is governed by an advection-diffusion equation and 
the corresponding MFPT results to be finite when the advecting velocity is in
the direction of the boundary. 
In particular, in the present study, 
we derive a nonhomogeneous 
Wiener--Hopf integral equation \cite{integralequations}
for the determination of the MFPT that allows 
indeed for its exact calculation and then for avoiding asymptotic limits.
This formula emerges to depend on the whole distribution of the jump-sizes and
on the mean-value only of the waiting-times. 
Thus, it holds for general non-Markovian random walks 
since, in CTRW, Markovianity intended as a local 
governing equation of the particle distribution 
(in opposition to integral equation in time) emerges solely with 
an exponential distribution of the waiting-times,
see, \cite{zwanzig1983classical} and discussion
around equation (2.6) in \cite{mainardi2000fractional}.
In this setting, 
we consider the quite general case study of a family of  
asymmetric distributions of the jump-sizes that is exponential
towards the boundary and arbitrary in the opposite direction.
By using the derived nonhomogeneous Wiener--Hopf equation, 
we can show that the MFPT is indeed independent of the 
jump-sizes distribution in the opposite direction to the boundary and 
this finding leads to the generalization of existing results as,
for example, 
those based on an asymmetric double-exponential jump-sizes distribution
\cite{gutkowiczkrusin_etal-jsp-1978,kou_etal-aap-2003}.
Moreover, we show that there exists a length-scale distinguishing
near-boundary and far-boundary starting points that depends only
on the features of the distribution of jumps in the direction of
the boundary.
Therefore, if for initial positions far-away from the boundary
a universal behaviour is observed, 
namely the MFPT does not depend on the specific distribution
of the jump-sizes, 
for initial positions near the boundary the emerging length-scale
is indeed dependent on the specific distribution of the jump-sizes
towards the boundary and then the universality is lost. 

The advection-diffusion framework applies strictly 
to the case when the arrival location is fixed, but
it can be assumed also as approximation when the mobility of the
walker, e.g., the predator, is much greater than the mobility of
the endpoint, e.g., the pray 
\cite{mckenzie_etal-bmb-2009,kurella_etal-bmb-2015}. 
Actually, such approximation for animal movement works in practice, 
for example, 
for modelling fish population \cite{faugeras_etal-jtb-2007} 
or when red foxes preying on duck nests \cite{sovada_etal-jwm-1995} 
and, in general, it is the best strategy for a prey to survive 
when the motion is on a lattice 
\cite{moreau_etal-pre-2003,moreau_etal-pre-2004}.

However, beside searching processes in the advection-diffusion setting, 
we want to highlight that the derived formalism and results 
play a role indeed also in the recent studies on stochastic resetting (SR) 
\cite{evans_etal-prl-2011,evans_etal-jpa-2011,evans_etal-jpa-2020} 
and in particular in its generalisation as
random amplitude stochastic resetting (RASR) \cite{dahlenburg_etal-pre-2021}.
SR refers to diffusive processes that are interrupted 
by a step-back to the origin while RASR is its generalisation 
in the sense that the step-back to the origin is replaced indeed
by a step-back with random amplitude. 
Recently, SR has been experimentally investigated
\cite{talfriedman_etal-jpcl-2020}, too.
Actually, a relation exists between processes with resetting
and optimal search strategies,
in particular with certain intermittent search processes 
\cite{evans_etal-prl-2011,evans_etal-jpa-2011}.

A remarkable feature of resetting events on the diffusive dynamics
is that non-stationary processes turn into stationary and
the corresponding MFPT from infinite turns into finite
\cite{evans_etal-jpa-2020,masopuigdellosas_etal-pre-2019}. 
Therefore, for such processes, 
it is of high interest the derivation of a formalism
for the determination of the survival probability 
as well as of the MFPT 
\cite{dahlenburg_etal-pre-2021,pal_etal-jpa-2016,
masopuigdellosas_etal-pre-2019}. 
To this aim, 
on the basis of the so-called {\it first renewal picture},
an approach was introduced in the framework of the SR
for determining the survival probability of Brownian diffusion 
in the presence of resetting interval lengths between subsequent 
resetting events that are exponentially distributed 
with time-dependent resetting rate \cite{pal_etal-jpa-2016}.
The same approach was later extended to more general random motions 
and arbitrary distribution of resetting interval lengths
\cite{masopuigdellosas_etal-pre-2019}. 

A similar approach for determining the survival probability 
as well as the MFPT in the case of RASR \cite{dahlenburg_etal-pre-2021}
is desired and not available yet. 
We report that RASR can be divided into dependent RASR 
and independent RASR and that
the simplest process of independent RASR is the CTRW.
Therefore, the Wiener--Hopf equation here derived and
the finding of a critical length-scale is a first-step in this
direction. Moreover, we report that 
the CTRW under resetting was considered for studying
the MFPT in the frame of anomalous diffusion with restarts 
\cite{bodrova_etal-pre-2020,mendez_etal-pre-2021}.

The reminder of the paper is organised as follows.
In the next Section \ref{sec:finiteMFPT} 
we provide the mathematical setting for a general random walk
that in the asymptotic limit is governed by an advection-diffusion
equation and 
we discuss how such a system displays a finite MFPT. 
Then we derive in Section \ref{sec:WH} the Wiener--Hopf 
equation for the exact computation of the MFPT and in
Section \ref{sec:casestudy} we put in action this formula
for the case study  
of a quite general family of asymmetric distributions 
of the jump-sizes that is exponential towards the boundary 
and arbitrary in the opposite direction.
Conclusions are finally reported in Section
\ref{sec:conclusions}. 

\section{Finite-mean first-passage time}
\label{sec:finiteMFPT}
Let $\mathbb{R}$ be the set of real numbers, 
we denote by $\mathbb{R}^+$ and by $\mathbb{R}_0^+$
the set of positive real numbers 
and the set of non-negative real numbers, i.e.,
$\mathbb{R}^+ = \{x \in\mathbb{R}|x > 0\}$ and 
$\mathbb{R}_0^+ = \{t \in \mathbb{R}| t \ge 0\}$, 
analogously we denote by $\mathbb{R}^-$ and 
by $\mathbb{R}_0^-$ the set of negative real numbers 
and the set of non-positive real numbers.

We consider a one-dimensional random walk in continuous-space $x \in \R$
and continuous-time $t \in \R_0^+$.
In particular, we study a CTRW characterised 
by a jump-sizes distribution $q : \R \to \R^+_0$ and  
by a waiting-times distribution
between consecutive jumps $\psi :\mathbb{R}_0^+ \to \mathbb{R}_0^+$, 
which normalise according to 
$\displaystyle{\int_{\R} q(\xi) \, \rmd \xi=1}$ and
$\displaystyle{\int_{\R^+_0} \psi(\tau) \, \rmd \tau=1}$, respectively.
If the distribution of the waiting-times is exponential then the
process is Markovian.
We assume that drawings of the jump-sizes and 
of the waiting-times are statistically independent, 
and both are independent and identically distributed (iid) random variables. 

Moreover, as it is reported in the Introduction, 
the present research aims also to be a preliminary study for further
advancements in SR 
\cite{evans_etal-prl-2011,evans_etal-jpa-2011,evans_etal-jpa-2020} 
and, more in general, in RASR theory \cite{dahlenburg_etal-pre-2021},
then within those frameworks the walker's trajectory 
can be interpreted as follows.
Let $X_t : \mathbb{R}_0^+ \to \mathbb{R}$ be the walker's position
at time $t \ge 0$, and let the notation $X_{t-\tau}|y$ denoting the actual 
walker's position at time $t$ provided that 
the walker was previously in $y \in\mathbb{R}$ 
at $\tau \in [0,t]$, which is the duration of the first random waiting-time.
By using the idea of the {\it first renewal picture} 
\cite{dahlenburg_etal-pre-2021, pal_etal-jpa-2016},
the walker may stay until time $t$ in the initial position $x_0 \in \R$ 
with probability $\Psi(t)=1-\int_0^t \, \psi(\tau) \, \rmd \tau$
or, 
it may be at time $t$ in $X_{t-\tau}$ starting 
from the new initial-like position $x_0+\xi,\, \xi \in \mathbb{R}$,
with the complementary probability $\int_0^t \, \psi(\tau) \, \rmd \tau$,
where $\xi$ denotes a random jump amplitude that may occurs 
at the new initial datum $\tau\in[0,t]$. 
In formulae, for a statistically homogeneous process,
the conditional PDF $p(x,t;x_0)$ of the walker emerges to be  
\cite{zaburdaev_etal-rmp-2015}
\begin{equation}
p(x,t;x_0)=\Psi(t)\delta(x-x_0)
+\int_0^t \! \psi(\tau) 
\! \int_{-\infty}^\infty 
\! q(\xi) p(x-\xi,t-\tau;x_0) \, \rmd\xi \rmd\tau \,,
\label{deriv_PDF_4}
\end{equation}
or
\begin{equation}
p(x,t;x_0)=
\Psi(t)\delta(x-x_0)
+\int_0^t \! \psi(\tau) 
\! \int_{-\infty}^\infty 
\! q(\xi-x_0) p(x,t-\tau;\xi) \, \rmd\xi \rmd\tau \,.
\label{PDF}
\end{equation}
In equation (\ref{deriv_PDF_4}),
variable $\xi$ represents the size of the first jump, 
while in (\ref{PDF}) it represents 
the starting position immediately after the first jump.

We consider a CTRW with finite-mean waiting-times, i.e.,
\begin{equation}
0 < \langle \tau \rangle = \int_0^\infty \tau \, \psi(\tau) \, \rmd\tau < 
+\infty \,, 
\label{wt_stat}
\end{equation}
and finite-mean and finite-variance jump-sizes, i.e.,
\begin{equation}
-\infty < \langle \xi \rangle =
\int_{-\infty}^\infty \xi \, q(\xi) \, \rmd\xi < +\infty \,, 
\label{js_stata}
\end{equation}
\begin{equation}
0 < \langle \xi^2 \rangle = 
\int_{-\infty}^\infty \xi^2 \, q(\xi) \, \rmd\xi < +\infty \,. 
\label{js_statb}
\end{equation}
Therefore, the diffusive limit $t \to \infty$ of the process
can be obtained in the corresponding limit $s \to 0$ of 
$\widetilde{\psi}(s)$, i.e.,
$\widetilde{\psi}(s)
\simeq 1- s \langle\tau\rangle + \mathcal{O}(s^2)$,
and, analogously, 
the tails $|x| \to +\infty$ 
of the walker's PDF in the corresponding limit $\kappa \to 0$ of
$\widehat{q}(\kappa)$, i.e.,
$\widehat{q}(\kappa)
\simeq 
1 + i\kappa\langle\xi\rangle - \kappa^2\langle\xi^2\rangle/2
+\mathcal{O}\lk k^3 \rk$,
where $\widetilde{\psi}(s)$ and
$\widehat{q}(\kappa)$ are the Laplace transform of $\psi(t)$ and
the Fourier transform of $q(x)$, respectively.
Thus the marginal distribution, namely 
$\displaystyle{P(x,t)= \int_{-\infty}^\infty p(x-y,t) P_0(y) \, \rmd y}$, 
with $P(x,0)=P_0(x)$, results to be governed by
the advection-diffusion equation 
\cite{redner-2001} 
\be
\frac{\partial P(x,t)}{\partial t}
= -v \, \frac{\partial P(x,t)}{\partial x}
+ D \, \frac{\partial^2 P(x,t)}{\partial x^2} \,, 
\label{advec_diff}
\ee
where
\be
v=\frac{\langle\xi\rangle}{\langle \tau \rangle} \,, \quad
D=\frac{\langle\xi^2\rangle}{2 \, \langle \tau \rangle} \,.
\label{Diff_const}
\ee
The solution of (\ref{advec_diff}) is 
\cite[page 16, formula (1.3.30)]{redner-2001}
\begin{equation}
P(x,t)= \int_{-\infty}^{\infty}
\frac{P_0(y)}{\sqrt{4\pi Dt}}\exp\lkk-\frac{(x-y-vt)^2}{4Dt} \rkk \, 
\rmd y \,.
\label{sol_adv_diff}
\end{equation}

Without loss in generality,
we restrict the initial position in $x_0 \in \Omega \subseteq \R^+$ 
and locate the target in $x=0$.
This means that the starting position of the
walker can not be located on the boundary.
Let $\lambda:\Omega\times\mathbb{R}_0^{+}\to\mathbb{R}^+$ 
be PDF of the first arrival at the absorbing boundary in $x=0$ 
at time $t \in \R_0^+$ for a given initial position in $x_0 \in \Omega$, 
then the survival probability,
namely the probability that the walker is not absorbed, 
up to the time $t$ is 
\be
\Lambda(x_0,t) = 1 - \int_0^t \lambda(x_0,\tau) \, \rmd\tau \,,
\quad t\in \mathbb{R}_0^+ \,, \quad x_0\in\Omega \subseteq \mathbb{R}^+ \,.
\label{surv_first_arriv}
\ee
Actually, the distribution $\lambda(x_0,t)$ and the conditional PDF
$p(x,t;x_0)$ are related. In particular,
if we assume that a walker, with initial position $x_0$, is arrived in $x=0$ 
at some instant $t-\tau$, with $\tau\in[0,t]$,
then $\lambda(x_0,t)$ provides the probability 
for the walker to return in $x=0$ after a time-interval $\tau$, 
namely at time $t$, 
and this gives the convolution integral 
\cite{montroll-1964,montroll_etal-jmp-1965,redner-2001}
\be
p(0,t;x_0)=\int_0^t \lambda(x_0,\tau) p(0,t-\tau;0) \, \rmd\tau \,.
\label{fatd}
\end{equation}
Formula (\ref{fatd}) can be derived also by 
including a sink $-\lambda(x_0,t)\delta(x)$ 
in the advection-diffusion equation (\ref{advec_diff})
\cite{mendez_etal-pre-2021,chechkin_etal-jpa-2003}. 

To conclude, 
by combining (\ref{surv_first_arriv}) 
and (\ref{fatd}) in the Laplace domain and by using (\ref{sol_adv_diff}),
we have \cite{redner-2001}
\begin{eqnarray}
\widetilde\Lambda(x_0,s)
&=& \frac{1}{s}\left[
1-\frac{\widetilde{p}(0,s;x_0)}{\widetilde{p}(0,s;0)}\right]
\nonumber \\
&=&\frac{1}{s}\left[
1-\exp\left\{- \frac{[v + \sqrt{v^2+4Ds}] \, x_0}{2D}\right\}\right] \,.
\end{eqnarray}
Thus, by applying the L'H{\^o}pital's rule, 
the MFPT results to be
\be
T(x_0)=\lim_{s\to 0} \widetilde\Lambda(x_0,s)
=-\frac{x_0}{v} > 0 \,, 
\quad {\rm with} \quad 
\langle \xi \rangle < 0 \,,
\label{MFPT_ad_di}
\ee
by reminding from (\ref{Diff_const})
that ${\rm sgn} v ={\rm sgn} \langle \xi \rangle$.
Actually, the linear dependence of the MFPT 
on $x_0$, as stated in formula (\ref{MFPT_ad_di}), 
is indeed universal,
namely it is independent of any specific distribution
of the jump-sizes whenever the limit leading to (\ref{advec_diff}) 
is met.

\section{A nonhomogeneous Wiener--Hopf integral equation for the MFPT}
\label{sec:WH}
The finite MFPT (\ref{MFPT_ad_di}) derived in Section \ref{sec:finiteMFPT}
holds in the limits $t, |x| \to +\infty$, 
or analogously $s, \kappa \to 0$, in which the behaviour is 
governed by an advection-diffusion equation (\ref{advec_diff}). 
This behaviour can be observed only for initial 
position $x_0 \to +\infty$, namely far from the absorbing boundary. 
As a matter of fact,
such asymptotic result is independent of the
specific distributions of the waiting-times and of the jump-sizes,
provided that (\ref{wt_stat}) and (\ref{js_stata},\ref{js_statb}) hold.

An alternative definition of the survival probability is indeed
\be
\Lambda(x_0,t)=\int_0^\infty p_{\rm abs}(x,t;x_0) \, \rmd x \,,
\quad \forall \, t \in\mathbb{R}_0^+ \,, \quad 
x_0 \in\mathbb{R}^+ \,,
\ee
where $p_{\rm abs}(x,t;x_0)$ is
the conditional distribution with an absorbing boundary at $x=0$.
From equation (\ref{PDF}) we have 
\be
p_{\rm abs}(x,t;x_0)=
\Psi(t)\delta(x-x_0)
+\int_0^t \!\! \psi(\tau) \! \int_{0}^\infty 
\!\! q(\xi-x_0)p_{\rm abs}(x,t-\tau;\xi) \, \rmd\xi \rmd\tau \,,
\label{PDF_abs}
\end{equation}
such that it holds
\be
\Lambda(x_0,t)
=\Psi(t)+\int_0^t \psi(\tau)\int_{0}^\infty 
q(\xi-x_0) \Lambda(\xi,t-\tau) \, \rmd\xi \rmd\tau \,.
\label{surv_prob}
\ee
Hence,
by passing through the Laplace transform, we obtain
\be
\widetilde{\Lambda}(x_0,s)=
\widetilde\Psi(s)+\widetilde\psi(s)\int_{0}^\infty 
q(\xi-x_0) \widetilde\Lambda(\xi,s) \, \rmd\xi \,, \quad
\forall\, x_0\in\mathbb{R}^+ \,.
\label{surv_prob_lap}
\ee
By plugging (\ref{surv_first_arriv}) into 
(\ref{surv_prob_lap}), we observe that, for each 
waiting-times distribution with finite mean, the first 
passage time density $\lambda(x_0,t)$ is dependent on the 
specific waiting-times distribution. 
Through the limit $s \to 0$ in (\ref{surv_prob_lap}),
we have that the MFPT $T(x_0)$ is
\begin{eqnarray}
T(x_0)=\langle\tau\rangle+\int_{0}^\infty q(\xi-x_0) T(\xi) \, \rmd\xi \,,
\quad \forall\, x_0\in\mathbb{R}^+ \,,
\label{fo1}
\end{eqnarray}
which is a {\it nonhomogenous} Wiener--Hopf integral equation 
\cite{integralequations} whose kernel is a PDF
\cite{spitzer-1957,spitzer-1960}.
The condition $x_0 \neq 0$
allows for avoiding to step into distribution theory,
which requires the definition of the kernel in such a way 
that the equation holds
at the endpoints \cite[Chapter 8]{integralequations}.
An inhomogeneous Fredholm equation for the MFPT was
indeed derived under different assumptions by using an
unbiased random walk for modelling 
animal movement and searching \cite{mckenzie_etal-bmb-2009}.

We observe that 
for discrete time-steps, i.e.,
$t \mapsto n \in \mathbb{N}$, we have 
$\psi(\tau)=\delta(\tau-1) \mapsto \delta_{n \,1}$ 
and $\Psi(t)=\Theta(1-t) \mapsto \delta_{n \, 0}$, 
such that formula (\ref{surv_prob}) reduces
to the well-known 
homogeneous Wiener--Hopf integral equation for
Markovian processes
\cite[equation (10)]{bray_etal-ap-2013} 
\be
\Lambda(x_0,n)
=\int_{0}^\infty 
q(\xi-x_0) \Lambda(\xi,n-1) \, \rmd\xi\,,
\label{surv_prob_discrete_t}
\ee
with boundary condition 
$\Lambda(x_0,0)=1$, $\forall \, x_0\in\mathbb{R}^+$. 
Clearly, discrete-time random walks are a special case of CTRW.
In particular, we stress that while equation (\ref{surv_prob_discrete_t}) 
is a straightforward special case of (\ref{surv_prob}),
equation (\ref{surv_prob}) cannot be derived 
from (\ref{surv_prob_discrete_t}).  
In fact, the nonhomogeneous nature of formula (\ref{surv_prob}) 
embodied by the probability of no jump in $[0,t]$ is a feature
that does not belong to a discrete-time setting.
Formula (\ref{surv_prob}) can not be derived from formula 
(\ref{surv_prob_discrete_t}) and this
allows for a more detailed analysis of the MFPT 
with respect to the existing literature, as it discussed in the following.
Moreover, from (\ref{fo1}),
we observe that the MFPT $T(x_0)$ depends  
on the whole distribution of the jump-sizes, 
that in general is asymmetric, and 
on the mean-value only of the waiting-times, 
while it is independent of the whole distribution of these last.
From the independence of (\ref{fo1}) of
the waiting-times distribution, 
we have that (\ref{fo1}) holds for general 
non-Markovian CTRWs, 
in fact it is known that solely in the special case
of an exponential distribution of waiting-times 
a CTRW is Markovian 
\cite{zwanzig1983classical,mainardi2000fractional}.  

Equation (\ref{fo1}) can indeed be extended to the $d$-dimensional case.
As a matter of fact, formula (\ref{PDF}) can be extended to 
a CTRW in $d$-dimensions such that,
by defining $q:\mathbb{R}^d\to\mathbb{R}_0^+$, 
for all $t\in\mathbb{R}_0^+$ and 
$\mathbf{x} \,, \mathbf{x}_0\in\mathbb{R}^d$, we have that 
\begin{equation}
p(\mathbf{x},t; \mathbf{x}_0)=
\Psi(t)\delta^d(\mathbf{x}-\mathbf{x}_0)
+\int_0^t \! \psi(\tau) 
\! \int_{\mathbb{R}^d} 
\! q(\bm{\xi}-\mathbf{x}_0) p(\mathbf{x},t-\tau;\bm{\xi}) \, 
\rmd^d {\bm{\xi}} \, 
\rmd\tau \,,
\label{PDF_ddim}
\end{equation}
which, in presence of an absorbing boundaries, turns into
\be
\hspace{-1.0truecm}
p_{\rm abs}(\mathbf{x},t;\mathbf{x}_0)=
\Psi(t)\delta^d(\mathbf{x}-\mathbf{x}_0)
+\int_0^t \!\! \psi(\tau) \! \int_{{\mathbb{R}^{d+}}}
\!\! q(\bm{\xi}-\mathbf{x}_0)p_{\rm abs}(\mathbf{x},t-\tau;\bm{\xi}) 
\, \rmd^d{\bm{\xi}} \rmd\tau \,,
\label{PDF_abs_ddim}
\end{equation}
where ${\mathbb{R}^{d+}} = \{(x_1, ..., x_d)^T \in\mathbb{R}^d|
x_i > 0 \,, \forall \, i \in \{1,..., d\} \}$.
Finally, by repeating the same steps from (\ref{PDF}) to (\ref{fo1}),
we obtain the following $d$-dimensional Wiener--Hopf equation 
for computing the MFPT of a $d$-dimensional CTRW
\begin{eqnarray}
T(\mathbf{x}_0)=\langle\tau\rangle+\int_{\mathbb{R}^{d+}} 
q(\bm{\xi}-\mathbf{x}_0) T(\bm{\xi}) \, \rmd^d{\bm{\xi}} \,,
\quad \forall\, \mathbf{x}_0\in{\mathbb{R}^{d+}} \,.
\label{fo1_ddim}
\end{eqnarray}

\section{The role of the distributions of the jump-sizes 
towards and away from the boundary}
\label{sec:casestudy}
We introduce now dimensional 
spatial-quantities $\xi^D$ and $x_0^D$ through 
a length-scale $\ell$:
\be
\xi^D = \ell \, \xi \,, \quad x_0^D = \ell \, x_0 \,,
\label{nondim1}
\ee
such that, consistently with (\ref{Diff_const}), it holds 
\be
\langle \xi^D \rangle = \ell \, \langle \xi \rangle = \ell\,v\, 
\langle \tau \rangle =v^D \langle \tau \rangle\,.
\label{nondim2}
\ee

Moreover, we consider the following quite general family
of distributions of the jump-sizes:
\begin{eqnarray}
\label{gen_jump_dist}
\rho(\xi^D)=q(-\xi^D)=
\left\{
\begin{array}{l l}
\displaystyle{
\frac{(1-b)}{\ell}\exp\lk-\frac{\xi^D}{\ell}\rk} \,; 
& {\rm if} \quad \xi^D\in \mathbb{R}_0^+ \,,\\
\\
b\, \rho_{<}(\xi^D) \,; & {\rm if} \quad \xi^D\in\mathbb{R}^- \,,
\end{array}
\right.
\end{eqnarray} 
that generalizes the double-exponential distribution 
studied in literature 
\cite{gutkowiczkrusin_etal-jsp-1978,kou_etal-aap-2003}.
In (\ref{gen_jump_dist}),
the length-scale $\ell$ represents the mean value of the jump-sizes
toward the boundary and 
parameter $b$ denotes the probability to jump away 
from the absorbing boundary. Thus, we distinguish
between exponentially-distributed jumps in the direction 
of the absorbing boundary and arbitrary jump-distributions 
in the opposite direction, such that
\be
\!\!
\int_{-\infty}^0\, \rho_{<}(\xi^D) \,\rmd \xi^D=1 \,, \quad 
\int_{-\infty}^0 \, \xi^D \, \rho_{<}(\xi^D) \rmd \xi^D = 
-\langle \xi^+\rangle=-\frac{\ell}{a} \,, \quad a > 0 \,,
\label{def:rho_minor}
\ee
The above formula (\ref{def:rho_minor}) 
defines the parameter $a$ that is the absolute value 
of the ratio between the mean value of the jump-sizes towards the boundary,
i.e., $-\ell$, and the mean value of the jump-sizes 
in the opposite direction to the boundary, i.e., $\langle \xi^+\rangle$. 
Thus, if $a > 1$ the jump-sizes towards 
the boundary are on average larger than the jump-sizes away from 
the boundary and vice versa if $a<1$. 
Indeed, if $a=1$ the mean value of the jump-sizes is the same in
both directions.

\begin{figure}[h]
\begin{center}
\subfloat[][]{
\includegraphics[height=6cm,width=10cm]{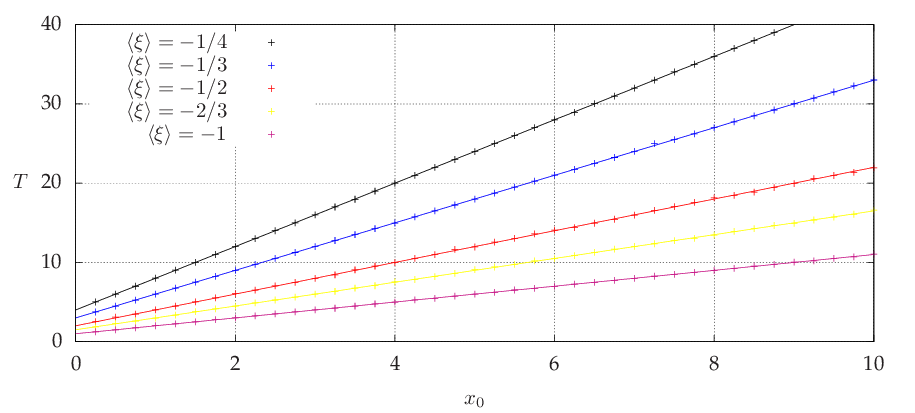}}
\\
\subfloat[][]{
\includegraphics[height=6cm,width=10cm]{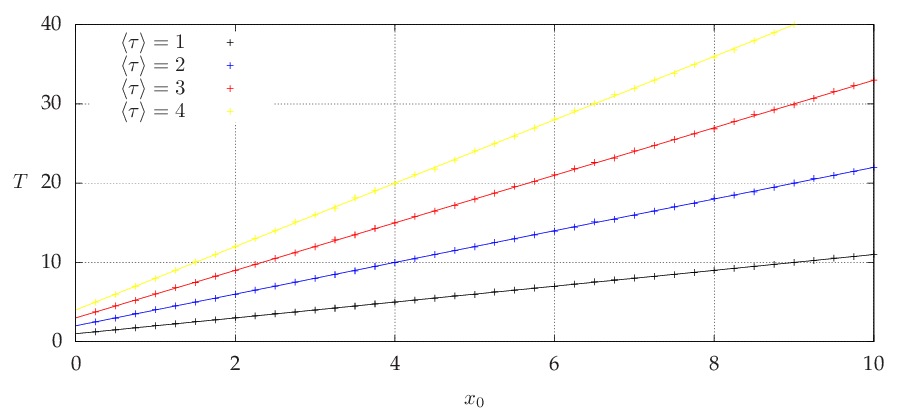}}
\\
\subfloat[][]{
\includegraphics[height=6cm,width=10cm]{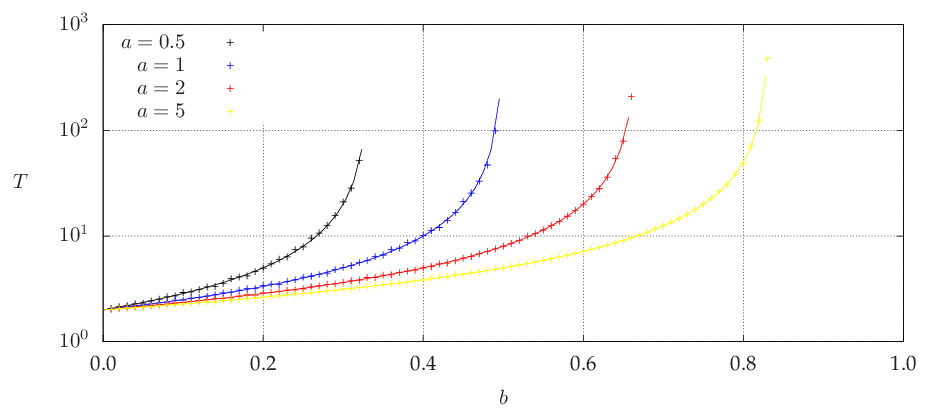}}
\end{center}
\caption{Comparison of the MFPT $T(x_0)$ computed by 
formula (\ref{gen_proof2}) (solid line)
and by the corresponding CTRW model (cross symbols).
Panel a) 
$T \, {\rm vs} \, x_0$ with $\langle\tau\rangle=1$ and varying $\langle\xi\rangle$;
Panel b)
$T \, {\rm vs} \, x_0$ with $\langle\xi\rangle=-1$ and varying $\langle\tau\rangle$;
Panel c)
$T \, {\rm vs} \, b$ with $x_0=\langle\tau\rangle=1$ and varying $a$.
}
\label{fig}
\end{figure}

For the average of the jump-sizes, we obtain
\begin{eqnarray}
\label{aver_jump_gen}
\langle\xi^D\rangle
&=&\int_{-\infty}^\infty\, \xi^D\,q(\xi^D) \rmd \xi^D
=-\int_{-\infty}^\infty\, \xi^D\,\rho(\xi^D)\, \rmd \xi^D \,,\nonumber\\
&=&-b\int_{-\infty}^0\, \xi^D\,\rho_{<}(\xi^D)\, \rmd \xi^D 
- \frac{(1-b)}{\ell} \int_0^{\infty} \, \xi^D \, 
\exp\lk-\frac{\xi^D}{\ell}\rk\,\rmd \xi^D \,, \nonumber\\
&=& b\,\langle \xi^+\rangle -(1-b)\,\ell
=\frac{b\,\ell}{a} -(1-b)\,\ell
=\ell \, \left[ \frac{1+a}{a} \, b - 1 \right] \,, \nonumber\\
&=&\ell\,\langle\xi\rangle \,, \quad
\langle \xi \rangle = \frac{1+a}{a} \, b - 1 \,, 
\end{eqnarray}
which is in agreement with formula (\ref{nondim2}) and,
for any arbitrary length-scale $\ell\in \mathbb{R}^+$, 
it fulfils the condition $\langle\xi^D\rangle<0$ if $a>b/(1-b)$  
or, equivalently, $b<a/(1+a)$, and we remark that
this last establishes also an upper bound for $b$ 
at which the MFPT becomes infinite, see figure \ref{fig}c.

In order to calculate the solution of (\ref{fo1}) with (\ref{gen_jump_dist}), 
in the spirit of the Wiener--Hopf technique 
\cite{abrahams_etal-siamjam-1990,leppington-1992,kisil_etal-prsa-2021}, 
we first generalise the MFPT to $T : \mathbb{R} \to \mathbb{R}$ and
equation (\ref{fo1}) reads
\begin{equation}
T(x_0)=f(x_0) + \int_0^\infty \rho(x_0-\xi) T(\xi) \, \rmd\xi \,, 
\quad x_0\in \mathbb{R} \,,
\label{fo4}
\end{equation}
where
\begin{eqnarray}\label{fo5}
T(x_0)=
\left\{
\begin{array}{l l}
T_{+}(x_0):\mathbb{R}^+\to\mathbb{R} \,, \\
\\
T_{-}(x_0):\mathbb{R}_0^-\to\mathbb{R} \,,
\end{array}
\right.
\end{eqnarray}
\begin{eqnarray}
\label{fo6}
f(x_0)=
\left\{
\begin{array}{l l}
f_{+}(x_0):\mathbb{R}^+\to\mathbb{R} \,, \\
\\
f_{-}(x_0):\mathbb{R}_0^-\to\mathbb{R} \,.
\end{array}
\right.
\end{eqnarray}

For our purposes, and without loss of generality, 
we set $f_{-}(x_0) = 0$, with
$x_0\in\mathbb{R}_0^{-}$, and then we have
\begin{equation}\label{fo7}
T_{-}(x_0)=\int_0^\infty \rho(x_0-\xi)T_{+}(\xi) \, \rmd\xi \,, 
\quad x_0\in \mathbb{R}_0^{-} \,.
\end{equation}
We introduce the generalised Fourier transform with 
$\kappa \in \mathbb{C}$
\begin{equation}
\widehat{T}(k)=\int_{-\infty}^\infty \exp(ikx_0)T(x_0) \, \rmd x_0 \,,
\label{fo8}
\end{equation}
that leads to the following pairs
\begin{eqnarray}
\widehat{T}_{\pm}(\kappa)
&=&{\pm}\int_{0}^{\pm\infty} 
\exp(+i\kappa x_0)T_{\pm}(x_0) \, \rmd x_0 \,, 
\label{fo9a} \\
T_{\pm}(x_0)
&=&\frac{1}{2\pi}\int_{L_{\pm}} 
\exp(-i\kappa x_0)\widehat{T}_{\pm}(\kappa) \, \rmd\kappa \,,
\label{fo9b} 
\end{eqnarray}
where $L_{\pm}$ are proper integration paths in the complex plane. 

By applying Fourier transform (\ref{fo8}) to equation (\ref{fo4})
and by using formulae (\ref{fo9a}) and (\ref{fo9b}), 
we observe that
$$
\widehat{T}(\kappa)=\widehat{T}_{-}(\kappa)+\widehat{T}_{+}(\kappa) \,,
$$
$$ 
\int_{-\infty}^\infty \exp(+i \kappa x_0)f(x_0) \, \rmd x_0
=\int_{0}^\infty \exp(+i \kappa x_0)f_+(x_0) \, \rmd x_0
=\widehat{f}_{+}(\kappa) \,,
$$
\begin{eqnarray*}
\label{fo11}
\fl
\int_{\mathbb{R}} \exp(+i \kappa x_0)\lkk 
\int_{\mathbb{R}^{+}} \rho(x_0-\xi)T_{+}(\xi) \, \rmd\xi \rkk \, \rmd x_0
&=& \int_{\mathbb{R}^{+}} \lkk \int_{\mathbb{R}} 
\exp(+i\kappa x_0) \rho(x_0-\xi) \, \rmd x_0 \rkk T_{+}(\xi) \, \rmd\xi \,,
\\
&=& \int_{\mathbb{R}^{+}} 
\exp(+i \kappa\xi) T_{+}(\xi) \, \rmd\xi 
\int_{\mathbb{R}} \exp(+i\kappa y) \rho(y) \, \rmd y \,,
\end{eqnarray*}
and we obtain
\begin{equation}\label{fo10}
\left[1-\widehat{\rho}(\kappa)\right] \widehat{T}_{+}(\kappa)
= - \widehat{T}_{-}(\kappa)+\widehat{f}_{+}(\kappa) \,.
\end{equation}
Moreover, since by comparing (\ref{fo1}) and (\ref{fo4})
we have that $f_{+}(x_0) = \langle \tau \rangle = const. > 0$ 
for $x_0\in\mathbb{R^+}$, then from the formula
\begin{eqnarray}\label{fo12}
\frac{1}{2\pi i}\int_{L_{+}} \frac{\exp(-i\kappa x_0)}{\kappa} \, \rmd\kappa=
{\rm Res} \, \frac{\exp(-i\kappa x_0)}{\kappa}=
\lim_{\kappa \to 0} \kappa \, \frac{\exp(-i\kappa x_0)}{\kappa}= 1 \,,
\end{eqnarray}
it holds
\begin{equation}
\hat{f}_+(\kappa) = - i \frac{\langle \tau\rangle}{\kappa} \,.
\label{fo13}
\end{equation}

By following the above procedure and by using the standard Wiener--Hopf method 
\cite{abrahams_etal-siamjam-1990,leppington-1992,kisil_etal-prsa-2021},
we have calculated the MFPT of CTRW models in continuous-space with
a paradigmatic asymmetric double-exponential distribution of 
jump-sizes and mean value $\langle \xi \rangle$ (see Appendix A).
The emerging non-uniqueness issues have been solved by imposing
constraints on physical consistency as: 
non-negativity, increasing monotonicity and the convergence 
to the expected asymptotic behaviour for starting-points 
far-away from the boundary.
Finally, the MFPT results to be 
\be
\label{fo31}
T_+ (x_0) = -
\frac{\langle\tau\rangle}{\langle\xi\rangle}(1+x_0) \,,
\quad \forall \, x_0\in\mathbb{R}^+ \,, 
\langle\xi\rangle\in \mathbb{R}^- \,,
\langle\tau\rangle\in \mathbb{R}^+ \,,
\ee
that is in agreement with the literature
\cite[page 513, Corollary 3.2]{kou_etal-aap-2003}.

By reversing the dimensional to non-dimensional relations
(\ref{nondim1}) and (\ref{nondim2}),
the ratio $x_0/\langle \xi \rangle$ in terms of non-dimensional quantities
becomes in dimensional quantities
\be 
\frac{x_0}{\langle \xi \rangle} =
\frac{x_0^D}{\ell} \frac{\ell}{v^D \langle \tau \rangle} = 
\frac{x_0^D}{v^D \langle \tau \rangle} \,, 
\quad 
\frac{1}{\langle \xi \rangle}= \frac{\ell}{\langle \xi^D \rangle} = \frac{\ell}{v^D\, \langle \tau\rangle} \,,
\ee 
and, in dimensional form,  formula (\ref{fo31}) reads
\be
T_+(x^D_0|\langle \xi^D\rangle) = - \frac{1}{v^D} (\ell+x^D_0) \,,
\quad \forall \,x^D_0\in\mathbb{R}^+ \,,  
v^D \in \mathbb{R}^-\,,  \ell \in \mathbb{R}^+ \,.
\label{solution_dim}
\ee
We remind that, 
the same asymmetric exponential-distribution of jumps 
that we considered here 
(\ref{fo3}) as a special case of (\ref{gen_jump_dist}),
it was considered joined with a diffusion process and
a constant drift with an exponential waiting-times distribution
\cite{kou_etal-aap-2003}, 
while a different family of asymmetric exponential-distribution of jumps
was indeed used in a partially related study on a one-dimensional lattice
\cite{gutkowiczkrusin_etal-jsp-1978}.

We show now that the MFPT of a CTRW with exponentially distributed
jumps towards the boundary is independent of the distribution of
the jumps in the opposite direction. 
In particular, if through the Wiener--Hopf integral equation (\ref{fo4}) 
we have that formula (\ref{solution_dim}) is the MFPT for
an asymmetric double-exponential distribution, 
we can show, again through (\ref{fo4}), 
that it is indeed also the solution 
when the jump-sizes are distributed according to (\ref{gen_jump_dist}).\\
Equation (\ref{fo4}) in dimensional form is 
\be
\label{gen_proof}
T_+(x^D_0|\langle\xi^D\rangle)
= \langle\tau\rangle + 
\int_0^\infty \rho(x^D_0-\xi^D) T_+(\xi^D|\langle\xi^D\rangle) \, \rmd\xi^D \,, 
\quad x^D_0\in \mathbb{R}^+ \,,
\ee
and by plugging (\ref{solution_dim}) into (\ref{gen_proof}) we have
\begin{eqnarray}
-\frac{1} {v^D}(\ell+x^D_0)
&=& \langle\tau\rangle - \frac{1-b}{v^D \, \ell}
\int_0^{x_0^D} \exp\lk \frac{\xi^D-x_0^D}{\ell}\rk (\ell+\xi^D) \, \rmd\xi^D 
\nonumber \\ 
& & \qquad \qquad 
- \frac{b}{v^D}\int_{-\infty}^0 \rho_<(\xi^D) (\ell+x_0^D-\xi^D) 
\, \rmd\xi^D \,, \nonumber\\
&=&
\langle\tau\rangle - \frac{(1-b)\,x_0^D}{v^D} 
- \frac{b\,(\ell+x_0^D+\langle\xi^+\rangle)}{v^D} \,, \nonumber\\
&=&
\frac{b\,\langle \xi^+\rangle -(1-b)\,\ell}{v^D} - 
\frac{x_0^D}{v^D} - \frac{b\,(\ell+\langle\xi^+\rangle)}{v^D} \,,
\nonumber\\
&=& - \frac{1}{v^D}(\ell+x^D_0) \,, \quad x^D_0\in \mathbb{R}^+ \,.
\label{gen_proof2}
\end{eqnarray} 
Thus, the MFPT of an asymmetric CTRW in continuous-space 
with exponentially distributed jumps in the direction of the boundary
is indeed independent 
of the distribution of the jumps in the opposite direction.
This result generalizes those based on the double-exponential
jump distribution, see, e.g., 
\cite{gutkowiczkrusin_etal-jsp-1978,kou_etal-aap-2003},
and it is new in literature.

The proof \pageref{gen_proof2}
has been checked against numerical simulations
of the corresponding CTRW models for different
jump-size distributions, see Appendix B for details. 
The plots of the comparisons are shown in figure \ref{fig} where, 
beside the linear growing of the MFPT $T(x_0)$ with respect to $x_0$
with varying $\langle\xi\rangle$ or varying $\langle\tau\rangle$,
the dependence of the MFPT on the involved parameters 
$a$ and $b$ is also displayed.

From formula (\ref{solution_dim}), 
we have that the asymptotic universal behaviour 
(\ref{MFPT_ad_di}) is attained when $x_0^D$ is much more far from the boundary 
than the length-scale $\ell$ and this last depends only on the features
of the jump-size distribution towards the boundary
(\ref{gen_jump_dist}, \ref{gen_proof}-\ref{gen_proof2}),
i.e.,
\be
T_+ (x_0^D) \sim -\frac{x_0^D}{v^D}\,,  \quad
v^D \in \mathbb{R}^- \,,
\quad \forall \, x_0^D \gg \ell \,,
\ee
with $0 \le b < a/(1+a)$ and $a \in \mathbb{R}^+$. 

Moreover, from (\ref{solution_dim}-\ref{gen_proof2}) 
in the setting (\ref{gen_jump_dist}),
we have that 
there exists a self-similarity property between the jump-sizes
and the initial position because the process scales 
through the same length-scale $\ell$ with respect to both.
By introducing a time-scale of the MFPT that 
is $\tau_{\rm MFPT}=-\ell/v^D=-1/v \in \R^+$, 
then the MFPT scales according to
\be
T_+ (x_0) = \tau_{\rm MFPT} \, 
\mathcal{T}_+\!\!\left(\frac{x_0^D}{\ell}\right) \,,
\quad \mathcal{T}_+(z)=1+z \,, 
\quad \forall \, x^D_0 \,, \ell \,, \tau_{\rm MFPT} \in\mathbb{R}^+ \,. 
\label{self_sim}
\ee
Hence, 
from formula (\ref{self_sim}) we may see that if the jump-sizes $\xi^D$ 
scales with respect to a larger (smaller) length-scale $\ell$ then 
also the resulting MFPT with an initial position $x_0^D$ scales 
with respect to a larger (smaller) length-scale.  

\section{Conclusions}
\label{sec:conclusions}
We studied the problem of a finite MFPT when the diffusion is ruled by 
a CTRW model characterised by waiting-times with finite mean 
and by jump-sizes with both finite mean and finite variance.

In particular, we obtain a nonhomogeneous
Wiener--Hopf integral equation (\ref{fo1}) that
allows for an exact calculation of the MFPT by avoiding asymptotic limits.
This formula results to depend on the whole distribution 
of the jump-sizes and on the mean-value only of the waiting-times,
thus it holds for general non-Markovian CTRWs. 
The derived Wiener--Hopf integral equation (\ref{fo1}) has
been used for the paradigmatic case of an asymmetric double-exponential
distribution of the jump-sizes and also for a more
general family of asymmetric distributions
of the jump-sizes, namely exponential towards the boundary 
and arbitrary in the opposite direction, and 
we derived two main results that are: 
$i)$ 
when the jumps towards the boundary are exponentially distributed 
then the MFPT is indeed independent 
of the jumps distribution in the opposite direction
and $ii)$ a length-scale emerges,
which depends only on the features of the distribution of jump-sizes 
in the direction of the boundary, 
that establishes a
criterion for distinguishing when the starting point is near and 
when it is far-away from the boundary. 
As a matter of fact, 
in opposition to the universal MFPT for starting points that are
far-away from the boundary,  
for starting points that are near the boundary this universality is lost,
specifically because of the dependence of such emerging length-scale 
on the specific distribution of the jump-sizes towards the boundary.
A scaling-law for the solution also emerges. 

Moreover, the derived Wiener--Hopf equation is supported  
by the comparison of the MFPT as calculated by using formula (\ref{fo1}) 
against the MFPT obtained as output of the simulations
of the corresponding CTRW models with different jump-sizes distributions. 

These findings can be viewed as an extension,
when the two processes are comparable,
of the results derived by Kou and Wang \cite{kou_etal-aap-2003},
who discussed a more general jump diffusion process
including both a Brownian motion and jumps,
together with a constant drift, but 
with an asymmetric double-exponential distribution of jump-sizes
and in the Markovian setting
by adopting exponentially-distributed waiting times.

To conclude,
we observe that, in the considered case (\ref{gen_jump_dist}),
the limit of the exact MFPT for initial positions approaching
the boundary is not zero but determined by the emerged length-scale,       
which is a parameter of the jump-sizes distribution
towards the boundary. Hence, 
this non-zero limit
provides also an indirect estimation of
the jump distribution towards the boundary as exponential
when the MFPT is known, for example, 
from data or from molecular simulations.

Concerning the application of the derived result, 
a finite MFPT is proper of 
some searching models that lead to advection-diffusion equations 
\cite{mckenzie_etal-bmb-2009,kurella_etal-bmb-2015}
but, beside this, we would like to highlight that a finite MFPT is 
indeed proper also of diffusive processes with stochastic resetting 
\cite{evans_etal-prl-2011,evans_etal-jpa-2011,evans_etal-jpa-2020}.
In this respect, we report that the 
Wiener--Hopf equation here derived and the corresponding approach
constitute a first step for determining the survival probability 
as well as the MFPT in the generalised stochastic resetting RASR 
\cite{dahlenburg_etal-pre-2021}, which are not available yet.
Therefore, an extension of the present approach for fulfilling
this purpose embodies a future perspective if this research.

To conclude, we would like to remind that
in spite of the fact that random walks, 
or at least their classical settings, seem to be fully understood,
some general features are still under investigation. 
In this respect, 
we have in mind the analysis 
concerning the fact that diffusive models meet
the Galilean invariance, in the best cases, 
solely weakly \cite{cairoli_etal-pnas-2018} and also the proof   
that exponential tails of walkers' PDF are indeed 
a universal property of diffusing particles at finite time, 
as well as at short time \cite{barkai_etal-prl-2020}.
The result here derived, despite obtained in a classical setting 
for the CTRW approach, joins with those lasts.

\ack 
This research is supported by the Basque Government through the 
BERC 2018--2021 program; 
by the Spanish Ministry of Science, Innovation and Universities through the 
BCAM Severo Ochoa excellence accreditation SEV-2017-0718 
and through the project PID2019-107685RB-I00 
and the Predoc Severo Ochoa 2018 grant PRE2018-084427.
The authors acknowledge the three anonymous referees for
their useful and constructive criticism that guided us to an
improvement of the research.

\appendix
\section{} 
We analyse the paradigmatic case study of a double-exponential 
distribution for the jump-sizes: 
\begin{eqnarray}
q(\xi)=
\left\{
\begin{array}{l l}
(1-b)\exp(\xi) \,; & {\rm if} \quad \xi\in \mathbb{R}_0^- \,,\\
\\
a b\exp(-a\,\xi) \,; & {\rm if} \quad \xi\in\mathbb{R}^+ \,,
\end{array}
\right.
\label{fo3}
\end{eqnarray}
where $b\in[0,1]$, $a \in \mathbb{R}^+$ and,
for lightening the notation, the length-scale $\ell$ 
has been dropped for a while.
The mean value of the non-dimensional $\xi$ results to be 
\begin{equation}
\langle \xi \rangle =\lk\frac{1+a}{a}\rk b - 1 \,,
\label{fo3a}
\end{equation}
that is consistent with (\ref{aver_jump_gen}).

\begin{eqnarray}
\label{fo6a}
\rho(\xi)=q(-\xi)=
\left\{
\begin{array}{l l}
(1-b)\exp(-\xi) \,; & {\rm if} \quad \xi\in \mathbb{R}_0^+ \,,\\
\\
ab\exp(a\,\xi) \,; & {\rm if} \quad \xi\in\mathbb{R}^- \,.
\end{array}
\right.
\end{eqnarray}

Since in the considered case (\ref{fo3}) 
it results $\hat{\rho}(k)=ab/(a+ik) + (1-b)/(1-ik)$, 
formula (\ref{fo10}) reduces to
\begin{equation}\label{fo14}
\widehat{T}_+(\kappa)=\widehat{T}_+^0 (\kappa) 
- i\langle\tau\rangle
\lk \frac{a+i\kappa(1-a)+\kappa^2}{\kappa^3+i\kappa^2(ab+b-a)} \rk \,,
\end{equation}
where $\widehat{T}_+^0 (\kappa)$ 
is the solution of the homogeneous case, i.e., $f_+(x_0) = 0$.
Solution $\widehat{T}_+^0 (\kappa)$ 
can be determined, by definition, up to a multiplicative constant 
(see formula (\ref{fo1}) with $\langle \tau \rangle = 0$) 
that here we denote by $C$:
\begin{equation}\label{fo15}
\widehat{T}_+^0 (\kappa)=
- C \, \frac{(a+i\kappa)(1-i\kappa)}{\kappa^2+i\kappa(ab+b-a)} 
\, \widehat{T}_{-} (\kappa) \,.
\end{equation}
By remembering the definition of $T_{-}(x_0)$ in (\ref{fo7}), 
that holds for $x\in\mathbb{R}_0^{-}$, 
we have that
\begin{eqnarray}\label{fo16}
\widehat{T}_{-}(\kappa)
&=& - \int_0^{-\infty} \exp[+i\kappa x_0] \lkk 
\int_0^\infty ab\exp(a(x_0-\xi))T_{+}(\xi) \, \rmd\xi\rkk \, \rmd x_0 
\nonumber \\
&=& - \int_0^{-\infty} \exp(+i\kappa x_0+ax_0) \, \rmd x_0 
\,\int_0^\infty ab\exp(-a \xi)T_{+}(\xi) \, \rmd\xi \nonumber \\
&=& \frac{T_{-}(0)}{a+i\kappa} \,,
\end{eqnarray}
and then
\begin{eqnarray}
\label{fo17}
\hspace{-1.3truecm}
T_+^0 (x_0)
= \frac{1}{2\pi}\int_{L_{-}} \exp(-i\kappa x_0)\hat{T}_+^0 (\kappa) \, 
\rmd \kappa 
\nonumber \\
= - C \, \frac{T_{-} (0)}{2\pi i}\int_{L_{-}} 
\frac{(i+\kappa)\exp(-i\kappa x_0)}{\kappa^2+i\kappa(ab+b-a)} \, \rmd\kappa
\nonumber \\
= - C \, T_{-} (0)\lkk 
\frac{1}{ab+b-a} 
+ \frac{(1+a-ab-b) \exp[(a-ab-b)x_0]}{a-ab-b} \rkk \,,
\end{eqnarray}
by remembering that
\begin{eqnarray}
\fl
\label{fo18}
{\rm Res}\lkk 
\frac{(i+\kappa) \exp(-i\kappa x_0)}{\kappa^2+i\kappa(ab+b-a)} \rkk 
&=& \lim_{\kappa \to 0} \lkk 
\kappa \, \frac{(i+\kappa)\exp(-i\kappa x_0)}{\kappa^2+i\kappa(ab+b-a)}\rkk
+ \nonumber \\
& & 
\lim_{k\to i(a-ab-b)} \lkk [\kappa+i(ab+b-a)] \, 
\frac{(i+\kappa)\exp(-i\kappa x_0)}{\kappa^2+i\kappa(ab+b-a)}\rkk 
\nonumber \\
&=& \frac{1}{ab+b-a} + \frac{(1+a-ab-b) \exp((a-ab-b)x_0)}{a-ab-b} \,.
\end{eqnarray}
Hence, by applying anti-transformation (\ref{fo9a}) to (\ref{fo14}) 
and by using (\ref{fo17}), we obtain 
\begin{eqnarray}\label{fo19}
\hspace{-2.0truecm}
T_+(x_0)
= -C \,T_{-} (0)\lk 
\frac{1}{ab+b-a}+\frac{(1+a-ab-b) \exp((a-ab-b)x_0)}{a-ab-b} \rk 
\nonumber \\
\qquad \qquad \qquad 
+ \frac{\langle\tau\rangle}{2\pi i} 
\int_{L_{-}} \frac{(a+i\kappa(1-a)+\kappa^2)\exp(-i\kappa x_0)}
{\kappa^3+i\kappa^2(ab+b-a)} \, \rmd\kappa \,,
\end{eqnarray}
that, after computing
\begin{eqnarray}
\fl
\label{fo20}
{\rm Res}\lkk 
\frac{(a+i\kappa(1-a)+\kappa^2)\exp(-i\kappa x_0)}
{\kappa^3+i\kappa^2(ab+b-a)} \rkk = 
\lim_{k\to 0} \frac{\rmd}{\rmd\kappa}\lkk \kappa^2 \, 
\frac{(a+i\kappa(1-a)+\kappa^2)\exp(-i\kappa x_0)}
{\kappa^3+i\kappa^2(ab+b-a)}\rkk \nonumber \\
\qquad + \lim_{\kappa\to i(a-ab-b)} \lkk 
[\kappa+i(ab+b-a)] \, \frac{(a+i\kappa(1-a)+\kappa^2)
\exp(-i\kappa x_0)}{\kappa^3+i\kappa^2(ab+b-a)} \rkk \nonumber \\
=
\frac{a x_0(a-b-ab)-a^2 b + b +a^2}
{(a-ab-b)^2} \nonumber \\
\hspace{3.0truecm}
-\frac{(ab+b)(1+a-ab-b)\exp((a-ab-b)x_0)}{(a-ab-b)^2} \nonumber \\
=\frac{a x_0}{a-ab-b}+
\frac{a^2 + b -a^2 b}{(a-ab-b)^2} \nonumber \\
\hspace{3.0truecm} 
+ \frac{(ab+b)(1+a)(b-1) \exp((a-ab-b)x_0)}{(a-ab-b)^2} \,,
\end{eqnarray}
becomes 
\begin{eqnarray}
\label{fo21}
\hspace{-2.0truecm}
T_+(x_0) 
=
- C \, T_{-} (0)\lkk 
\frac{1}{ab+b-a}+\frac{(1+a-ab-b) \exp[(a-ab-b)x_0]}{a-ab-b} \rkk 
\nonumber \\
\qquad + \langle\tau\rangle\lkk 
\frac{a x_0}{a-ab-b}
+\frac{a^2 + b -a^2 b}{(a-ab-b)^2} \right. \nonumber \\
\hspace{3.5truecm} \left.
+\frac{(ab+b)(1+a)(b-1) \exp((a-ab-b)x_0)}{(a-ab-b)^2} \rkk \,.
\end{eqnarray}
Moreover, constant $C$ can be estimated by calculating $T_{-}(0)$ 
through definition (\ref{fo7}) by using (\ref{fo21}) and,  
by remembering that
$$
\int_0^\infty \exp(-a\xi) \, \rmd\xi = \frac{1}{a} \,,
\quad 
\int_0^\infty \xi\exp(-a\xi) \, \rmd\xi =\frac{1}{a^2} \,,
\quad a > 0 \,,
$$
we obtain $C = -1$. 
Finally, the desired solution is
\begin{eqnarray}
\label{fo23}
\hspace{-2.0truecm}
T_+(x_0)
=T_{-} (0)\lkk 
\frac{1}{ab+b-a}+\frac{(1+a-ab-b) \exp((a-ab-b)x_0)}{a-ab-b} \rkk 
\nonumber \\
\quad +
\langle\tau\rangle\lkk 
\frac{a x_0}{a-ab-b}+\frac{a^2 + b -a^2 b}{(a-ab-b)^2} \right. \nonumber \\
\hspace{3.5truecm} \left.
+\frac{[(ab+b)(1+a)(b-1)] \exp[(a-ab-b)x_0]}{(a-ab-b)^2} \rkk \,,
\end{eqnarray}
that, by using (\ref{fo3a}), can be written in terms of $\langle\xi\rangle$:
\begin{eqnarray}
\label{fo23aaa}
\hspace{-2.0truecm}
T_+(x_0)
= T_{-} (0)\lkk 
\frac{1}{a\langle\xi\rangle}-\frac{(1-a\langle\xi\rangle) 
\exp(-a\langle\xi\rangle x_0)}{a\langle \xi \rangle} \rkk \nonumber \\
\quad + 
\langle\tau\rangle\lkk 
-\frac{x_0}{\langle\xi\rangle}
+\frac{\langle\xi\rangle(1-a)+1}{a\langle\xi\rangle^2} \right. \nonumber \\
\hspace{3.5truecm} \left.
-\frac{(\langle\xi\rangle+1)(1-a\langle\xi\rangle) 
\exp(-a\langle\xi\rangle x_0)}{a\langle\xi\rangle^2} \rkk \,.
\end{eqnarray}
In the case $\langle\xi\rangle=0$, 
formula (\ref{fo23aaa}) reduces to
\be\label{fo23aab}
T_+(x_0; \langle\xi\rangle= 0)
=T_{-} (0)\lk x_0+1 \rk + \langle\tau\rangle
\lkk 1+(1-a)x_0-\frac{a\,x_0^2}{2} \rkk \,,
\ee
which can be derived by using the series expansion of the exponential function,
and in the symmetric case $a=1$, from (\ref{fo23aab}) it results 
\be
T_+(x_0; \langle\xi\rangle= 0, a=1)
= T_{-} (0)\lk 1+x_0 \rk + \langle\tau\rangle
\lkk 1-\frac{x_0^2}{2} \rkk \,,
\label{fo23ab}
\ee
that are both (\ref{fo23aab}) and (\ref{fo23ab}) not MFPT solutions.

In fact, any MFPT solution has to fulfil, by definition, the conditions
\be
\frac{\rmd T_+}{\rmd x_0} > 0 \,, \quad \forall \, x_0\in\mathbb{R}^+ \,,
\label{fo26a}
\ee
\be
T_+(x_0) \ge 0 \,, \quad \forall \, x_0\in\mathbb{R}^+ \,.
\label{fo26b}
\ee
Therefore, for solution (\ref{fo23aaa}),
condition (\ref{fo26a}) is fulfilled if
\begin{eqnarray}\label{fo27}
\frac{dT_+(x_0)}{dx_0} = 
T_{-} (0) (1-a\langle\xi\rangle) 
\exp(-a\langle\xi\rangle x_0) \nonumber \\
+ \langle\tau\rangle \lkk 
-\frac{1}{\langle\xi\rangle}
+\frac{(1+\langle\xi\rangle)(1-a\langle\xi\rangle)
\exp(-a\langle\xi\rangle x_0)}{\langle\xi\rangle} \rkk > 0 \,, \quad
\forall \, x_0\in\mathbb{R}^+ \,,
\end{eqnarray}
which implies
\be
T_{-} (0) > 
\langle\tau\rangle \left[ 
\frac{\exp(a\langle\xi\rangle x_0)}
{\langle\xi\rangle(1-a\langle\xi\rangle)}
-\frac{1+\langle\xi\rangle}{\langle\xi\rangle} \right] \,, \quad
\forall \, x_0\in\mathbb{R}^+ \,,
\ee
where $(1-a\langle\xi\rangle) \ge 0$ as it follows from (\ref{fo3a}).
Hence, since it holds that 
\begin{eqnarray}\label{fo27a}
\fl
\sup_{x_0\in\mathbb{R}^+}
\left\{
\langle\tau\rangle \left[
\frac{\exp(a\langle\xi\rangle x_0)}
{\langle\xi\rangle(1-a\langle\xi\rangle)}-
\frac{1+\langle\xi\rangle}{\langle\xi\rangle}\right] \right\} =\left\{
\begin{array}{l l}
\infty \,, & {\rm if } \quad \langle\xi\rangle\ge 0 \,, \\
\\
\displaystyle{
-\langle\tau\rangle\frac{1+\langle\xi\rangle}{\langle\xi\rangle} \,,} &
{\rm if} \quad \langle\xi\rangle < 0 \,,
\end{array}
\right. 
\end{eqnarray}
inequality (\ref{fo27}), and then condition (\ref{fo26a}), 
is fulfilled if 
\begin{eqnarray}\label{fo28}
T_{-} (0)=-\langle \tau \rangle\frac{1+\langle\xi\rangle}{\langle\xi\rangle}
+ c \,, \quad \forall \, c\in \mathbb{R}_0^+ \,,
\quad \langle\xi\rangle\in \mathbb{R}^{-} \,.
\end{eqnarray}
When formula (\ref{fo28}) is plugged into (\ref{fo23aaa}),
we have that for $x_0\in\mathbb{R}^+$, $c\in\mathbb{R}_0^+$,
$\langle\xi\rangle\in \mathbb{R}^-$,
\begin{equation}
\label{fo29}
T_+ (x_0) =
-\frac{\langle\tau\rangle}{\langle\xi\rangle}(1+x_0)
+ \frac{c}{a\langle\xi\rangle}
[ 1-(1-a\langle\xi\rangle) \exp(-a\langle\xi\rangle x_0) ] \,,
\end{equation}
which fulfils condition (\ref{fo26b}), too.
Regarding constant $c$, 
we observe that formula (\ref{fo29}) meets the asymptotic 
linear growing with respect to $x_0$ (\ref{MFPT_ad_di}) 
when it holds $c=0$.
To conclude, the MFPT of CTRW models in continuous-space is 
\be
T_+ (x_0) = -
\frac{\langle\tau\rangle}{\langle\xi\rangle}(1+x_0) \,,
\quad \forall \, x_0\in\mathbb{R}^+ \,, 
\langle\xi\rangle\in \mathbb{R}^- \,,
\langle\tau\rangle\in \mathbb{R}^+ \,,
\ee
that is formula (\ref{fo31}).

\section{} 
In figure \ref{fig},
formula (\ref{gen_proof2}) is tested against the corresponding CTRW model.
In particular, the trajectories of the CTRW are generated 
by the iterative procedure
\begin{eqnarray}
\label{iteration}
\left\{
\begin{array}{l l l l}
X_n=X_{n-1}+\xi_n \,, 
\quad X_0=x_0 > 0 \,, \quad n \in\mathbb{N} \,,\\
\\
t_n=t_{n-1}+\tau_n \,, \quad t_0=0 \,, \quad n\in\mathbb{N} \,,
\end{array}
\right.
\end{eqnarray} 
with the random jump-sizes drawn according 
to distribution (\ref{gen_jump_dist}) with $\ell=1$,
i.e.,  
\begin{eqnarray}
\label{iteration_rv}
\xi_n=\left\{
\begin{array}{l l}
\ln(1-u^{(1)}_1)\,, & {\rm if} \quad u_2>b\,, \\
\\
\displaystyle{\frac{\chi_m}{m}} \,, 
\quad m \in[1,...,10] \,, & {\rm otherwise} \,, 
\end{array}
\right.
\end{eqnarray} 
where $\chi_m$ is a Poisson distributed random variable with mean $m/a$,
\begin{eqnarray}
\label{iteration_pois}
\chi_m=\chi_{m-1}\,-\,\displaystyle{\frac{\ln(1-u^{(m)}_1)}{a}} \,, 
\quad \chi_0= 0 \,, \quad m \in[1,...,10] \,,
\end{eqnarray}
and the random waiting-times are drawn from one of the following 
three distributions
\be
\tau_n = -\langle\tau\rangle \ln(1-u_3) \,,
\quad {\rm exponential} \, {\rm distribution} \,,  
\ee
\be
\tau_n= 2 \langle\tau\rangle \, u_4 \,, \quad  
{\rm uniform} \, {\rm distribution} \,, 
\ee  
\be
\tau_n= \langle\tau\rangle \,, \quad  
{\rm delta} \, {\rm distribution} \,, 
\ee  
such that 
$u^{(j)}_1 \,, u_2 \,, u_3 \,, 
u_4 \sim U(0,1) \,, 
\forall \, j\in[1,...,m]$, 
$b\in[0,1]$ and $a\in\mathbb{R}^+$.
We obtain the same results with all the distributions of waiting-times 
and all the distributions of jump-sizes 
in the opposite direction to the boundary 
as expected from formula (\ref{gen_proof2}).

The absorbing boundary located in $x=0$
can be passed only by a jump-event and, 
thus, the first passage-time (FPT) can be numerically computed by 
\begin{equation}
\label{FPT_numerics}
{\rm FPT} = t_n \,, 
\,\, {\rm provided} \, {\rm that} \,\,
X_n < 0 \,\, {\rm and} \,\, X_i >0 \,,
\quad \forall i = 1,...,n-1 \,.
\end{equation}
The MFPT $T_+(x_0)$ is
\begin{equation}
\label{MFPT_numerics}
T_+(x_0) = \frac{1}{N} \sum_{j=1}^N \, {\rm FPT}_j \,, 
\quad N\in \mathbb{N} \,,
\end{equation}
where $N$ is the number of independent realizations of the iteration procedure
(\ref{iteration}).

\section*{References}
\bibliographystyle{unsrt}
\bibliography{all-mfpt-pre} 

\end{document}